\documentclass[journal,onecolumn,11pt]{IEEEtran}

\input packages-headers

\RequirePackage[top=3.0cm,bottom=3.0cm,left=2.5cm,right=2.5cm,nohead]{geometry}

\newcommand{\MATLAB}{MATLAB$^\text{\textregistered}$}

\begin{document}
\renewcommand{\refname}{References}

\pagestyle{plain}
\thispagestyle{empty}

\title{Achieving Phase-based Logic Bit Storage in Mechanical Metronomes}

\author
{
Tianshi Wang\\
{\small
Department of Electrical Engineering and Computer Sciences, University of
California, Berkeley, CA, USA}\\
Email: \texttt{tianshi@berkeley.edu}
\vspace{-1em}
}

\maketitle

\begin{abstract}
Recently, oscillator-based Boolean computation has been proposed for its
potentials in noise immunity and energy efficiency.
In such a system, logic bits are encoded in the relative phases of oscillating
signals and stored in injection-locked oscillators.
To show that the scheme is very general and not specific to electronic
oscillators, in this paper, we report our work on storing a phase-based logic
bit in the relative phase between two mechanical metronomes.
While the synchronization of metronomes is a classic example showing the
effects of injection locking, our work takes it one step further by
demonstrating the bistable phase in sub-harmonically injection-locked
metronomes --- a key mechanism for oscillator-based Boolean computation.
Although we do not expect to make computers with metronomes, our study
demonstrates the generality of this new computation paradigm and may inspire
its practical implementations in various fields, \eg, MEMS, silicon photonics,
spintronics, synthetic biology, \etc{}

\end{abstract}

\thispagestyle{empty}
\section{\normalfont {\Large Introduction}}\seclabel{intro}

Perhaps the most pressing problem in the progress of electronics and
computation is the slowdown of Moore's law
\cite{MO1965mooreslaw,Wilson2013ITRS}.
Among the many design and manufacturing roadblocks, noise and variability are
having an ever-greater impact on system performance as transistors are further
miniaturized \cite{Wilson2013ITRS}; power consumption has also been a serious concern for many years
\cite{Esmaeilzadeh2011darkSilicon}.
While broad directions being explored to address them focus mainly on new
devices and new architectures
\cite{hisamoto2000finfet,Shulaker2013CNTcomputer,BehinDatta2010allSpinLogic},
recent research on oscillator-based Boolean computation
\cite{WaRoUCNC2014PHLOGON,RoPHLOGONprocIEEE2015,WaRoOscIsing2017} provides a
new perspective: alternative physical representation of logic bits.
Instead of using voltage levels, logic values 0 and 1 can be encoded in the
relative phase difference of an oscillating signal from a reference signal
(\figref{logic01}).
Such phase-based encoding features inherently superior noise immunity over
level-based encoding; a quick analogy is the use of phase and frequency
modulation over amplitude modulation in radio communication for better
resistance to noise and interference \cite{rappaport1996wireless}.
This advantage of phase encoding, although having been known in communication
for a long time, is yet to be fully exploited in computation.

\begin{figure}[htbp]
	\centering
	{
		\epsfig{file=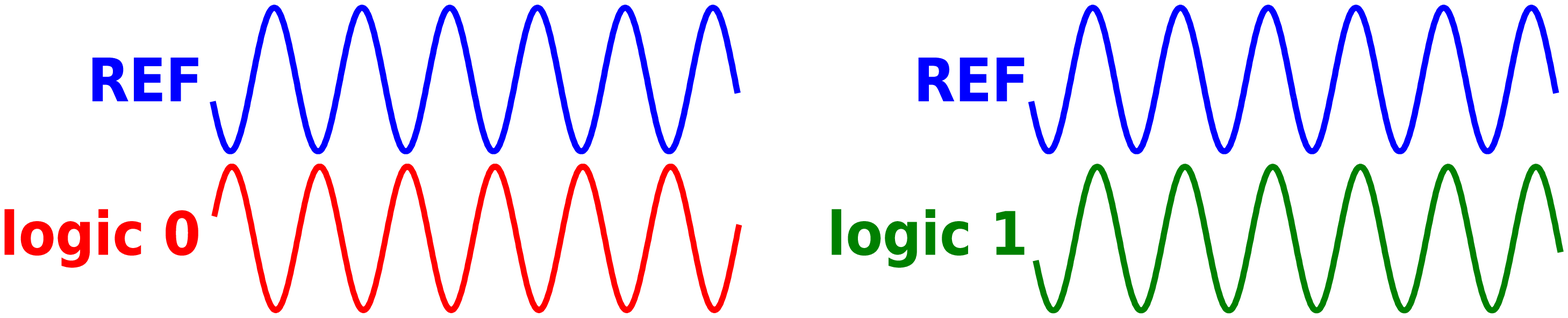, width=0.4\linewidth}
	}
    \caption{Encoding logic values in the phases of oscillating signals.\figlabel{logic01}}
\end{figure}

A key attractiveness of this oscillator-based computation scheme is its
generality --- almost any nonlinear self-sustaining oscillator can be used as a
logic latch.
Such an oscillator can be from any physical domain, including electrical (\eg,
using CMOS), biological (neurons, intracellular oscillators), nanotechnological
(Spin Torque Nano-Oscillators, MEMS resonators), optical (lasers), \etc; there
is great potential for high-speed and low-power operation
\cite{WaRoUCNC2014PHLOGON,RoPHLOGONprocIEEE2015}.
This generality comes from a mechanism known as injection locking (IL), which
is observed in almost all oscillators.
Under IL, an oscillator's response locks on to an oscillating perturbation in
both frequency and phase.
When the perturbation oscillates at integer multiples of the oscillator's
natural frequency, the oscillator's phase can lock to the input with
multistable sub-harmonic phase-locked responses.
This variant of IL is known as Sub-harmonic Injection Locking (SHIL).
The sub-harmonic responses can then be used to encode and store logic bits,
making the oscillator behave as a logic latch \cite{WaRoUCNC2014PHLOGON}.
For example, a binary latch can be implemented by perturbing an oscillator with
a small periodic signal at about twice of its natural frequency.
The oscillator will develop one of the two bistable responses with a phase
difference of $180^\circ$, representing 0 or 1 in phase encoding.
Together with phase-based combinational logic gates \cite{WaRoUCNC2014PHLOGON},
finite state machines can then be implemented for general-purpose Boolean
computation.

IL is ubiquitous in nature, \eg, the synchronization of fireflies' patterns,
neurons firing in unison, \etc; it is also widely used in engineering.
One famous example of IL is the demonstration of metronomes synchronizing their
ticks.
As illustrated in \figref{metronome-sync}, metronomes placed on a common
rolling platform receive small perturbations from their neighbours, and
eventually lock to the same frequency with the same phase, despite having
slightly different central frequencies and random initial phases.

\begin{figure}[htbp]
\centering{
    \epsfig{file=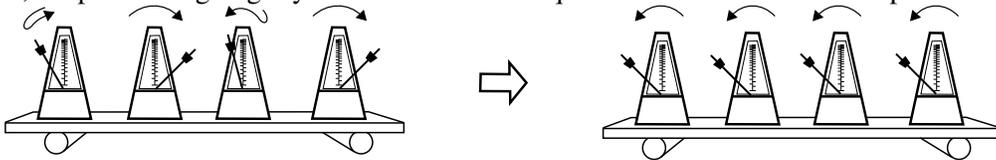,width=0.8\linewidth}
}
\caption{Metronomes standing on a rolling board end up ticking in unison.
    \figlabel{metronome-sync}}
\end{figure}

The setup in \figref{metronome-sync} can also be adapted to demonstrate SHIL
\cite{Wa2017arXivMetronomes}.
As is shown in \figref{SHIL-Lissajous}, on a common platform sit two metronomes
--- one oscillates at approximately the $1/2$ sub-harmonic of the other.
The pendulum tips of the metronomes have colored markers taped to them so that
their coordinates can be recorded through processing the video of oscillation.
We plot the x coordinates of one metronome's tip with respect to the other in
\figref{SHIL-Lissajous}; the resulting curves are known as the Lissajous curves.
When the platform is stationary, the Lissajous curves span the whole space,
indicating that their phases drift apart;
when the platform is rolling, providing injection between the two metronomes,
through the mechanism of SHIL their swing patterns synchronize.

\begin{figure}[htbp]
\centering{
    \epsfig{file=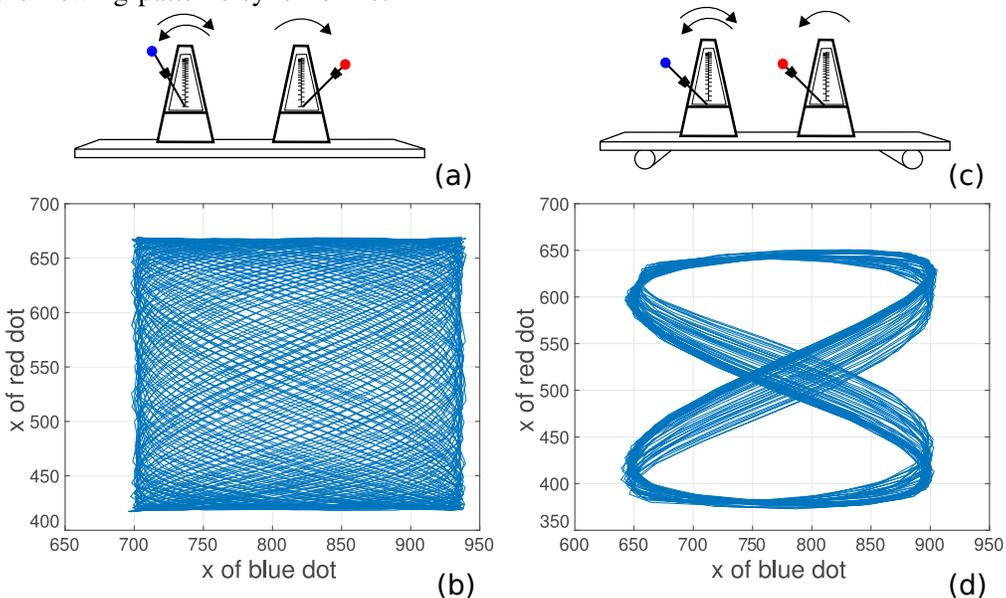,width=0.8\linewidth}
}
\caption{(a) Two metronomes tuned to approximately 1Hz and 2Hz are placed on a
stationery platform; (b) the corresponding Lissajous curves in show that no
synchronization happens; (c) the metronomes are coupled through a rolling
board; (d) the pattern in the Lissajous curves demonstrates SHIL.
    \figlabel{SHIL-Lissajous}}
\end{figure}

When SHIL happens, the oscillator with the sub-harmonic response features
bistable phase, allowing it to operate as a binary logic latch.
This has been demonstrated in CMOS ring oscillators and LC oscillators
\cite{WaRoUCNC2014PHLOGON,WaRoDAC2015MAPPforPHLOGON}.
Naturally, a demonstration with metronomes is desirable as it proves the
generality of the scheme.
For this purpose, the results from \figref{SHIL-Lissajous} are not enough, as
logic encoding based on phase requires a reference, \eg, another metronome
tuned to approximately 1Hz.
However, simply putting another 1Hz metronome on the same platform does not
work, as the two 1Hz metronomes will synchronize their phases --- the
bistability induced by SHIL from the 2Hz oscillator is not strong enough to
sustain a $180^\circ$ phase difference between them for encoding the other
logic value.
To make the two 1Hz metronomes reliably store a phase-based logic bit, one has
to be creative; the conventional setup for metronome experiments has to be
changed.
In \secref{results}, we explore a new experimental setup, where the two 1Hz
metronomes swing in directions perpendicular to each other, in order to avoid
injection locking between them.
Results from this setup are also analyzed in \secref{results}, proving that it
operates reliably as a binary logic latch that can store a phase-based bit.

All that the experimental setup requires are normal mechanical
metronomes\footnote{The metronomes used are Wittner Taktell Super-Mini
Metronome.} and some common household items such as tapes, a foam board and
several ping-pong balls; a cell phone is used for video recording and
processing the video requires minimal coding experience.
The demonstration is both interesting and easily reproducible.
As such, it has the potential to reach a broad audience and have a wide impact.

It is worth mentioning that phase-based logic storage has been demonstrated in
mechanical systems before.
A nanomechanical beam driven by AC power can develop sub-harmonic parametric
oscillation with a bistable phase \cite{Mahboob2008bit,Freeman2008nems}.
It has been considered as a first step towards nanomechanical computer
\cite{Freeman2008nems} using a parametric-oscillator-based
(``Parametron''-based) scheme
\cite{vonNeumann:1954:NLC:nonote,Wigington:1959:ProcIRE,Goto:1954,Goto1960patent}.
In comparison, the scheme we have discussed in this paper has several advantages
in the potential of miniaturization and system-level routing
\cite{WaRoUCNC2014PHLOGON} thanks to the use of DC-powered self-sustaining
oscillators.
To our knowledge, our work with the metronomes is the first to demonstrate bit
storage in the mechanical domain within this oscillator-based computation
paradigm.

Another similar work demonstrating phase-based logic bit storage is done
recently using chemical reactions \cite{gizynski2017chemical}.
It uses Belousov--Zhabotinsky reaction, which is a self-sustaining chemical
oscillator that creates periodically changing color.
In the experiments, three Belousov--Zhabotinsky droplets are coupled together.
They change color in a merry-go-round fashion; the two stable rotational modes
(clockwise and anticlockwise) represent logic value 0 and 1.
The demonstration creates the first ``one-bit chemical memory unit'' or the
``chit'' \cite{gizynski2017chemical}.
It is worth noting that storing information in clockwise and anticlockwise
rotations is a special case of phase-based encoding.
In fact, in our experimental setup, when the two 1Hz metronomes oscillate with
the two stable phase differences 0 and $180^\circ$, the platform they sit on
slightly rotates clockwise and anticlockwise respectively.
So our demonstration creates a similar one-bit mechanical memory unit with
self-sustaining oscillators, the first of its kind.

\section{\normalfont {\Large Experimental Setup and Results}}\seclabel{results}

As discussed in \secref{intro}, two metronomes tuned to approximately the same
frequency ($\approx$1Hz) are needed to represent a phase-based logic bit.
We also need another metronome tuned to about twice of their frequency
($\approx$2Hz) to provide them with a common injection.
Under SHIL, the two 1Hz metronomes will each develop a bistable phase; their
phase difference can then be used to store a logic bit.

However, we need to make the 2Hz metronome injection lock the two 1Hz ones
``independently'', \ie, the two 1Hz ones should not injection lock each other.
Towards this end, we design a metronome placement scheme different from the
conventional one in \figref{metronome-sync}.
The two 1Hz metronomes are placed with a $90^\circ$ angle, such that the
directions in which they swing are perpendicular.
To truly decouple them, the platform now needs to be able to roll freely in the
horizontal plane.
So instead of using cylinders as in \figref{metronome-sync}, we use balls to
support the platform on the table.
The third metronome, which is tuned to 2Hz, is then placed in between the two
1Hz ones, with a $45^\circ$ angle from both of them.
In this way, its swing injection locks them simultaneously.
The setup is illustrated in \figref{bit-storage-setup}.

\begin{figure}[htbp]
\centering{
	\begin{minipage}{0.5\linewidth}
      \epsfig{file=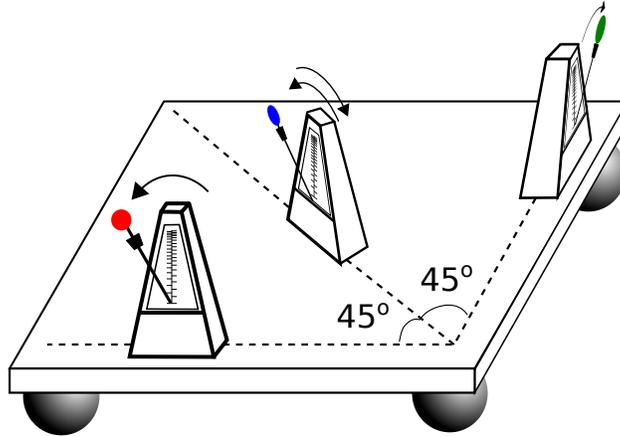,width=\linewidth}
	  \caption{Illustration of the experimental setup, where two 1Hz metronomes
		(with red and green tapes) oscillate in perpendicular directions, and
		one 2Hz metronome is placed between them with a $45^\circ$ angle.
		\figlabel{bit-storage-setup}}
	\end{minipage}
}
\end{figure}

There are many everyday objects that can be used as the platform and balls in
\figref{bit-storage-setup}.
In practice, we would like to minimize the total mass of the setup in order to
maximize the injection between metronomes.
Therefore, we use a small foam board instead of wooden ones, and four ping-pong
balls instead of marbles.
Such details can make a difference in the reliability of the setup and the
reproducibility of the results.

\begin{figure}[htbp]
\centering{
    \epsfig{file=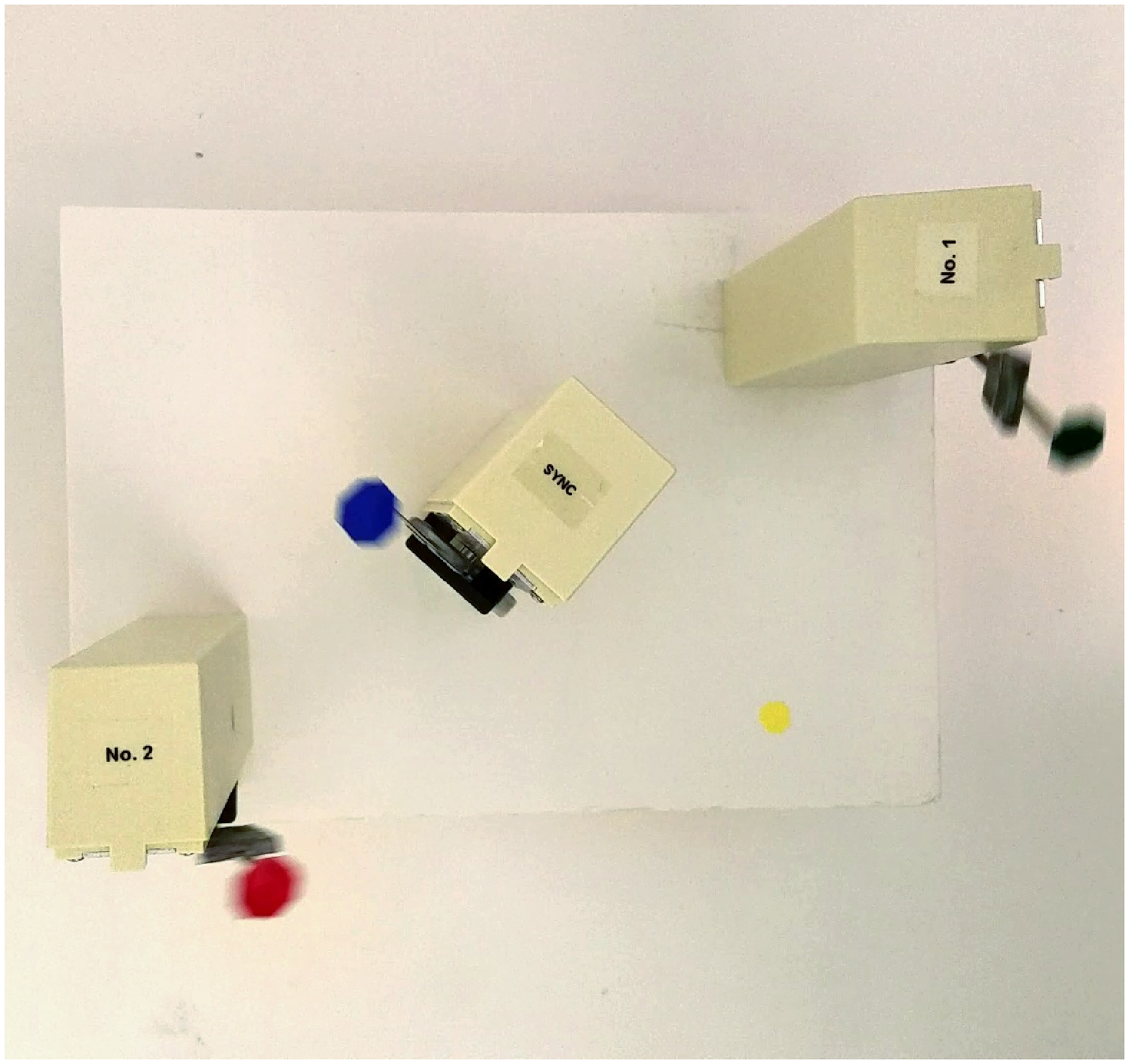,width=0.4\linewidth}
}
\caption{Photo of the experimental setup.
    \figlabel{metronome-photo}}
\end{figure}

As shown in the photo of the actual setup in \figref{metronome-photo},
the two 1Hz metronomes are taped with red and green markers, the 2Hz one is
taped with a blue marker.
We start the two 1Hz metronomes first and confirm that their ticking are not
synchronized due to frequency detuning.
Then we start the 2Hz one.
Through the mechanism of SHIL, the two 1Hz metronomes are both frequency locked
to the 2Hz one, and they develop a stable phase difference, which stores one
logic value.
We can manually stop one of the 1Hz metronomes for half a cycle then let it
resume its oscillation.
After the two 1Hz metronomes are synchronized by the 2Hz one again, 
their stable phase difference is changed from before by $180^\circ$,
representing the other binary logic value.
We record the whole process using a cell phone with a 60Hz frame rate.
The video is imported frame by frame into \MATLAB.
Then a simple algorithm is used to extract the locations of markers in each
frame.
\figref{bit-storage-waves} shows the oscillation of the coordinates of the
markers through the whole experiment of about 270 seconds.

\begin{figure}[htbp]
\centering{
    \epsfig{file=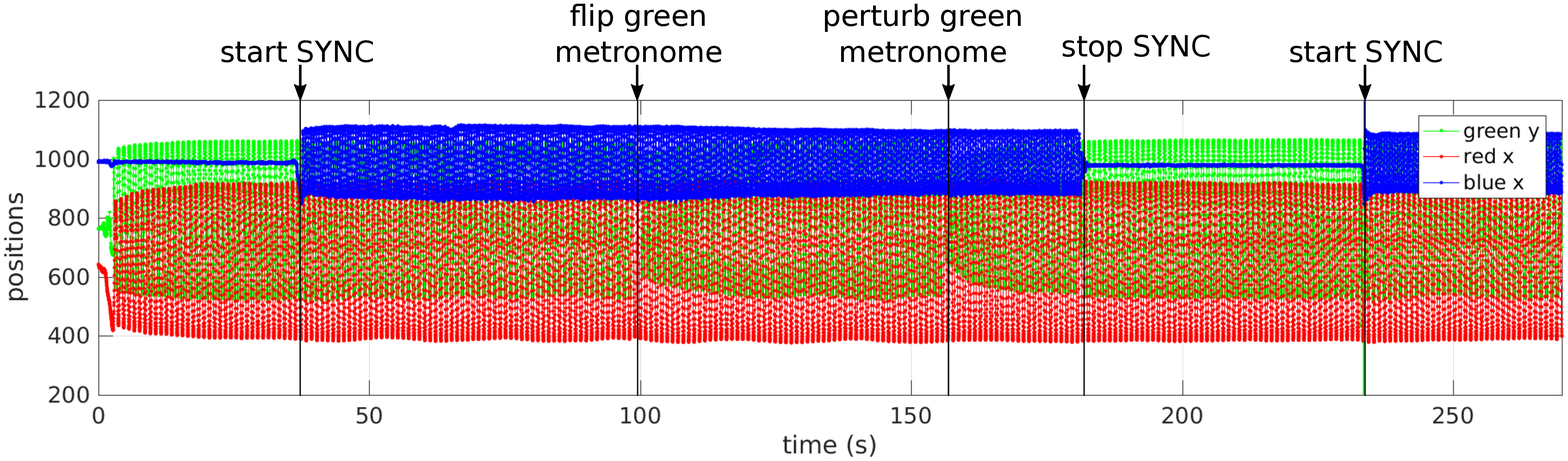,width=1.0\linewidth}
}
\caption{X or Y coordinates of the tips of the three metronomes in
	\figref{metronome-photo} during the experiment.
    \figlabel{bit-storage-waves}}
\end{figure}

\begin{figure}[htbp]
\centering{
    \epsfig{file=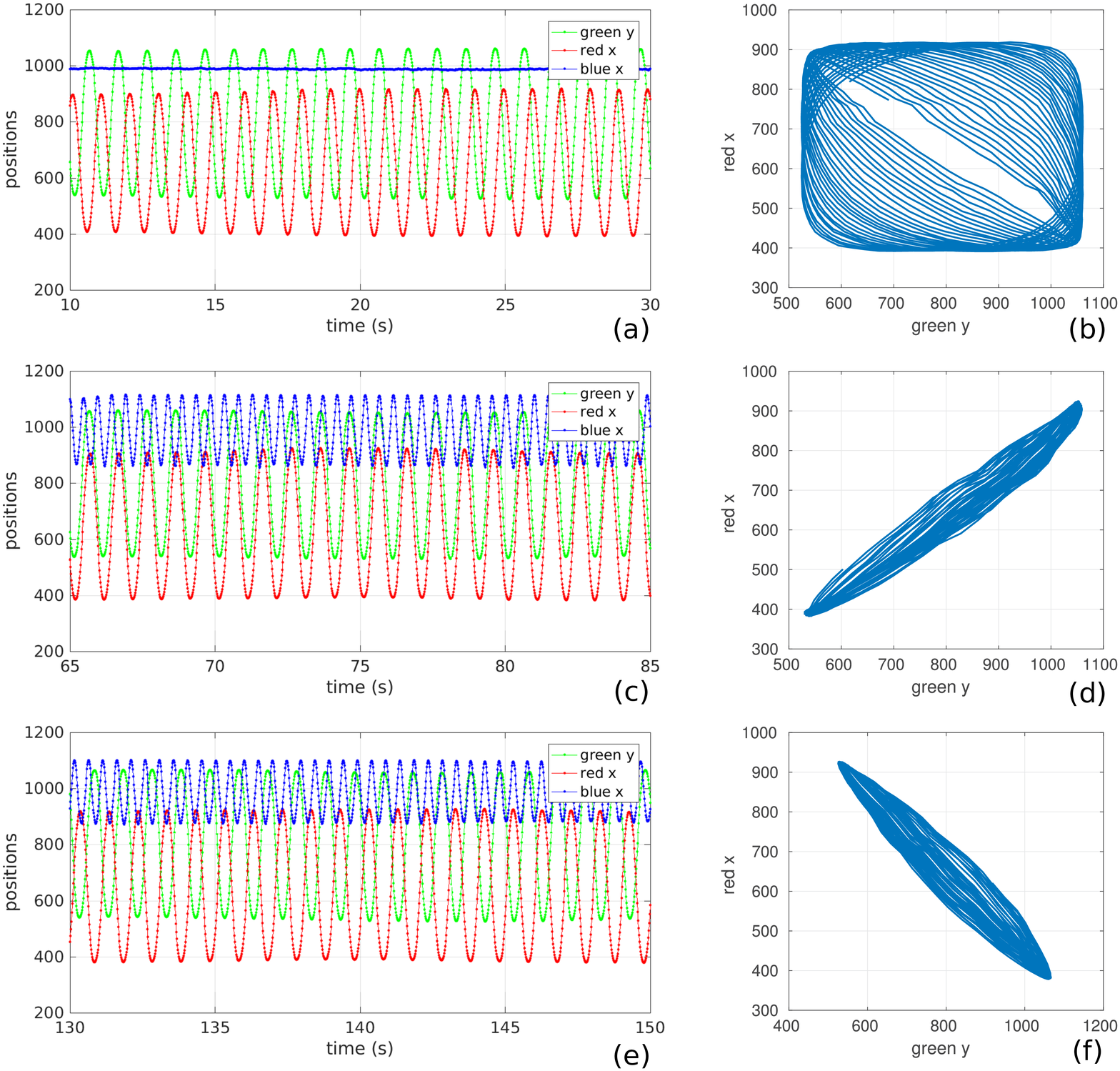,width=0.9\linewidth}
}
\caption{Excerpts from \figref{bit-storage-waves} and their corresponding
    Lissajous curves.
    \figlabel{bit-storage-waves-Lissajous}}
\end{figure}

In \figref{bit-storage-waves-Lissajous}, we show some excerpts from
\figref{bit-storage-waves}.
The first excerpt in \figref{bit-storage-waves-Lissajous} (a) shows the
oscillation of the two 1Hz metronomes from 10s to 30s, during which time the
2Hz one has not been started.
Careful observation indicates that the green and red waveforms are not
synchronized --- the red peak is aligned with the green valley at the beginning
of the time slot but has apparently become misaligned towards the end.
The corresponding Lissajous curves more clearly show that the phase difference
drifts with time.
In comparison, the second excerpt (\figref{bit-storage-waves-Lissajous} (c)) is
taken between time 65s and 85s, when SHIL is present.
We observe that the green and red peaks are well aligned, as can be confirmed
by the corresponding Lissajous curves in \figref{bit-storage-waves-Lissajous}
(d).
Similarly, after the green metronome's phase is flipped, from time 130s to
150s, the red peaks are aligned with the green valleys
(\figref{bit-storage-waves-Lissajous} (e)).
This other stable phase difference represents the other phase-encoded logic
value.
In these two stable states, the corresponding Lissajous curves in
\figref{bit-storage-waves-Lissajous} (d) and (f) both form a line, but the two
lines are perpendicular to each other.

To explore the stability of the stored bit, we perturb the green metronome at
about 155s in the experiment, as is shown in \figref{bit-storage-waves}.
Unlike bit flipping, we touch the metronome to slightly delay its oscillation.
As a result, the two 1Hz metronomes do not settle to the new phase difference;
their previous stable phase difference is restored within ten cycles.
This further validates that the setup is a bistable system storing a
phase-encoded bit.

\section{\normalfont {\Large Summary}} \seclabel{conclusion}

This paper explores the generality of the key mechanism of oscillator-based
Boolean computation by demonstrating bit storage in mechanical metronomes.
Through a creative setup, two sub-harmonically injection-locked metronomes have
been shown to develop a bistable phase difference of either 0 or $180^\circ$,
achieving a phase-encoded one-bit mechanical memory.
As injection locking is a ubiquitous phenomenon in all types of oscillators,
this demonstration has the potential of inspiring more implementations of
phase-encoded logic latches using oscillator technologies from various physical
domains.

\let\em=\it
\bibliographystyle{unsrt}
{\scriptsize\bibliography{stringdefs,tianshi,jr,PHLOGON-jr}}

\end{document}